\documentclass[letterpaper, paper,11pt]{AAS}	% for preprint proceedings
\usepackage{bm}
\usepackage{amsmath}
\usepackage{subfigure}
\usepackage[colorlinks=true, pdfstartview=FitV, linkcolor=black, citecolor= black, urlcolor= black]{hyperref}
\usepackage{overcite}
\usepackage{footnpag}	
\usepackage{amssymb}
\PaperNumber{25-400}

\begin{document}

\title{Increasing Determinism of All-Vs-All Orbital Evolution Simulators}

\author{Enrico M. Zucchelli\thanks{Postdoctoral Associate, Aeronautics \& Astronautics, Massachusetts Institute of Technology, 02139, MA.},  
Daniel Jang\thanks{Technical Staff, MIT Lincoln Laboratory, 224 Woods St. Lexington, 02420, MA,},
\ and Richard Linares\thanks{Associate Professor, Aeronautics \& Astronautics, Massachusetts Institute of Technology, 02139, MA.}
}

\maketitle{}

\begin{abstract}
All-vs-all orbital evolutionary simulations for the low Earth orbit~(LEO) simulate the long term evolution of the LEO environment. Although these simulations typically offer the highest fidelity, they are also highly computationally intensive. One factor that effectively reduces the efficiency of the approach is that all-vs-all approaches are stochastic and the distribution of the output has large variance. This paper introduces a new, quasi-deterministic all-vs-all simulator, whose variance is greatly reduced compared to traditional methods. The proposed approach virtually simulates collisions happening everywhere and all the time; however, their effect is appropriately reduced to maintain an unbiased estimate of the mean. Additional techniques are used to augment the proposed approach and obtain very precise estimates of any number of standard deviations from the mean, for the evaluation of the Value at Risk~(VaR) with a single, low-variance run.
Depending on the settings, results show that the variance in total number of debris generated can be reduced by a factor that averages 1,500, while increasing the computational cost by a factor of less than 1.5. Variance can be reduced even more when computing the VaR, albeit a no longer negligible bias is introduced. Low-variance results enable several key applications, such as sensitivity analysis, sustainability assessment of small missions, and fast evaluation of collision risk induced by existing debris. Additionally, rapid computations of the VaR can improve the evaluation of policy robustness, and include confidence intervals in risk assessment.
\end{abstract}

\section{Introduction}
The occupancy rate of the low Earth Orbit~(LEO) regime has been increasing rapidly in recent years. New trends, such as miniaturization of spacecraft and decreasing launch costs, have led to the proliferation of active satellites. The sudden growth in the number of artificial satellites launched in recent years has raised concerns about sustainability of the environment. Sustainability in space generally requires avoiding Kessler's syndrome~\cite{Kessler1978CollisionBelt}, a situation in which a runaway effect is triggered by collisions between anthropogenic objects in space. Evaluating if, when, and where the conditions for Kessler's syndrome start appearing requires simulations capable of predicting the future of the LEO environment. These simulations can then be used to define future policies~\cite{lifson2024low}, assess the risk of planned constellations and missions~\cite{Lucken2019CollisionDesign}, or define which defunct satellites to remove from orbit with most urgency~\cite{MITRI}. The most accurate predictors are all-vs-all simulators. In these simulations, all satellites are propagated forward in time; collisions are simulated and resulting debris are generated.

Accurate long term prediction of a collision event is a virtually impossible task. Even to date, the best collision estimates only offer probabilistic outcomes, and conjunctions are not even considered if they are set to happen more than a few weeks later. For this reason, most all-vs-all simulators rely on stochastically determining whether a collision between two satellites is bound to happen or not. Further, the resulting fragments are also generated randomly, as the state-of-the-art tool used is the NASA Standard Breakup Model~(SBM)~\cite{nasaSTMevolve4}, which computes a probabilistic distribution of fragments.
All-vs-all Monte Carlo methods include NASA's LEGEND~\cite{legend2004}, UKSA's DAMAGE~\cite{Lewis2001DAMAGE:Framework}, ESA's DELTA~\cite{DELTA2}, JAXA's LEODEEM~\cite{hanada2013orbital}, Aerospace's ADEPT~\cite{ADEPT2020}, MIT's open-source MOCAT-MC~\cite{Jang2024Monte}, and several others~\cite{medee,luca2,sharmaspace,iadccomparison}.
A single simulation with all-vs-all propagators has very large variance in the number of objects, even with short prediction spans: just one randomized collision can add up to 5,000 objects to the environment. The variance of all collisions sum up in a single simulation, leading to potentially very large final uncertainty.

The first contribution of this paper is to provide a theoretical justification for some of the settings required to use the cube method~\cite{LiouCUBE2003,LiouCUBE2006}, which is the approach of choice to compute probability of collisions for software such as MOCAT-MC and DAMAGE. While previous studies have established empirical guidelines on tuning time-step and cube size~\cite{LewisCUBElimitations2019,facchinetti2023analysis}, this work begins with a theoretical justification for those results. 
Then, the main contribution of this paper is to propose an alternative to Monte Carlo all-vs-all simulators: a quasi-deterministic all-vs-all evolutionary model whose variance is greatly reduced, requiring much fewer simulations than traditional Monte Carlo evolutionary models. Results show that the proposed approach, named MOCAT quasi-deterministic (MOCAT-QD) offers solutions that are almost unbiased with respect to the original corresponding all-vs-all simulator, and reduce the variance by a factor of up to 1,500, with computational cost only about 1.5 times larger. 
Finally, the proposed quasi deterministic simulator is augmented to be able to provide not only the mean, but also a user-defined $n_\sigma$ standard deviations from the mean, with a single run. The latter tool is necessary for robust planning, and for computing metrics such as the Value at Risk~(VaR).

The proposed methods rely on reducing the variance deriving from each source of randomness one by one. This work focuses on three main origins: 1) satellite collisions and resulting debris generation, 2) initial conditions, and 3) the success rate of post mission disposals~(PMDs). Uncertainties due to launch and rocket body explosions are currently neglected. The software presented in this paper makes the assumption that collisions are independent from one another; however, that is not true, since fragments generated by a collision can later on cause another collision. Nonetheless, we make the claim that the dependence is small, and can, in most cases, be neglected. This assumption is empirically validated for MOCAT-QD, but starts to show its limitations in MOCAT-QD-Var.
In this research we apply the proposed modifications to the Monte Carlo~(MC) version of MOCAT~\cite{
MOCAT-bin-AMOS2022,MOCAT-capacity-AMOS2022,mocat-ssem,rodriguez2024towards,MOCAT3}.
The majority of the proposed variance reduction techniques consist of averaging and rare event simulation~(RES)~\cite{lecuyer2009splitting,beck2015rare}. Examples of use of RES in astronautics include maneuvering target tracking~\cite{zucchelli2024bayesian} and recently, computation of probability of collision~\cite{SZTAMFATERGARCIA2025}. The crux of this paper lies in providing a method that exploits concepts from RES to estimate the expected number of fragments resulting from multiple rare collisions.
For stochastic nonlinear dynamics, averaging is known to introduce a bias in estimators such as the mean and covariance. The empirical results from this work provide upper bounds for the bias introduced by the novel approach; however, the lower bound is 0 for almost all tested cases.
Increased determinism in simulations not only provides higher computational efficiency, but also enables extensive sensitivity analysis as well as risk and sustainability assessments. %Further, the generation of more deterministic scenarios enables efficient data generation used for the training of MOCAT - machine learning~(MOCAT-ML), which is a deterministic ML model that promises to provide fast, high-fidelity predictions of the LEO environment~\cite{rodriguez2024towards}.

\section{All-vs-all Simulators}
All-vs-all simulators predict the state of the LEO environment by propagating the motion of all satellites in LEO. The main components are shortly described in this section. For a more detailed discussion on all-vs-all simulators, and specifically on MOCAT-MC, the software of choice for this research, the reader may refer to Jang et al.~\cite{Jang2024Monte}.
\subsection{Orbit Propagation}
In MOCAT-MC objects are propagated deterministically. The propagation utilizes an augmentation of Brouwer-Lyddane theory~\cite{brouwer1959solution,lyddane1963small,Martinusi2015} to allow for low inclination and small eccentricities. The propagator includes the effects of atmospheric drag as well as of the perturbations due to $J_2$. As the approach is analytical, the propagation error is independent of the time-step.
\subsection{Collisions}
Collisions are the largest variance-inducing component of all-vs-all simulators. In MOCAT-MC the probability of collision between any two objects is computed with the cube method~\cite{LiouCUBE2003,LiouCUBE2006}. Space is discretized in cubes of a preset size; after every integration step satellites are binned in the cubes. If any two objects are in the same cube, a probability of collision is computed assuming the space objects are two gas particles residing in the same cube for a duration of time equal to the integration step:
\begin{equation}
\label{eq:pc_cube}    p_{c,ij} = \frac{|\bm{v}_i-\bm{v}_j|\sigma_{i,j}} {V} \Delta t
\end{equation}
where $p_{c,ij}$ is the collision probability between objects $i$ and $j$, $\bm{v}_i$ is the velocity of object $i$, $\sigma_{i,j}$ is the cross-sectional collision area of the two objects, $V$ is the volume of the cube, and $\Delta t$ is the integration time-step.
Several studies have shown that the cube method provides on average realistic results as long as $\Delta t$ and $V$ are properly chosen, and MOCAT-MC has been validated against several other LEO evolutionary simulators.
The fragments resulting from a collision are generated according to NASA's SBM~\cite{nasaSTMevolve4}. If one very large object is at risk of colliding with an even moderately small one, the number of generated debris of size above 0.1~m can be larger than 5,000. Based on some preliminary analysis, there is approximately a 5\% chance every year (assuming the current orbital environment) of having a collision that produces more than 3,000 trackable debris. These events alone lead to a standard deviation of 500 objects in one year. As a result, MC runs have a standard deviations of about 1,500 objects after only 5 years of propagation.
% Need to double check the numbers for this one, the evaluation was very rough
\subsection{Deorbit}
Deorbiting is the process by which the LEO environment cleans itself up. Through drag, the semi-major axis of the orbit of a satellite is reduced. The effect of drag strongly depends on the solar cycle, and increases exponentially with lower altitudes. After a satellite's periapsis is below a certain altitude, the object is considered to have deorbited.
\subsection{Post-Mission Disposals}
An ever larger number of space missions include post-mission disposals~(PMDs) at the end of their lifecycle. PMD is a maneuver that decreases the energy of a satellite after the mission has been completed. PMD leads the satellite to deorbit earlier than it would have if the deorbiting were left to natural causes. The timing of the deorbiting subsequent to PMD is mission dependent, and ranges from a few minutes to many years. In MOCAT-MC the PMD is simulated as instantaneous. To include possible failures that will inevitably occur, a PMD maneuver is simulated to succeed with a probability less than 1, whose value is set by the user. For a success rate of 0.95, and assuming 2,000 PMD maneuvers a year, the standard deviation in total number of failed PMDs after 10 years is about 30. Although the number of these objects is small in relative terms, they are whole satellites, generally larger and more massive than debris; hence, a collision involving one of these objects would be much more consequential than a collision involving any two fragments. Further, the number of failed PMDs increases greatly for scenarios with large launch rates.

\section{Cube Size Selection}
\begin{figure}
    \centering
    \includegraphics[width=0.5\textwidth]{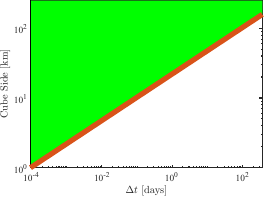}
    \caption{Regions of the cube side vs. time-step space that avoid bias for fixed cross-sectional diameter of 50~m (green).}
    \label{fig:dt_and_cube}
\end{figure}

There are two major factors in play when setting the default size of the cube: accuracy, or more appropriately, bias, and computational speed. The \textit{total} number of pairs evaluated grows approximately linearly with the volume of the cube, hence cubically with its side. Usually, the practitioner would want to be in a situation where most of the satellites do not share a cube with any other satellite. This would allow to maximally exploit the main advantage of the cube method, which is the reduction of the number of collisions to be considered.
However, too small a cube leads to a bias. The bias is introduced because the probability of collision is inversely proportional to the volume of the cube: a small volume may then lead the probability of collisions to be higher than 1. Any time that the probability of collision is larger than one, a bias is introduced in the simulation.
Computing the minimum cube volume $dV_{min}$ such that the probability of collision is always smaller than or equal to 1 is equivalent to the following equality:
\begin{equation}
\label{eq:unbiasedCube}
    1 = \frac{{\max_{i,j,t}|\bm{v}_{i,t}-\bm{v}_{j,t}|\,\sigma_{i,j,t}\, I^b_{i,j,t}}}{dV_{min}}\, \Delta t,
\end{equation}
where $I^b_{i,j,t}$ is the binning indicator function, equal to 1 if objects $i$ and $j$ are binned in the same cube at time $t$, and 0 otherwise. As Eq.~\eqref{eq:unbiasedCube} requires a full simulation to be evaluated, simplifications can be made to obtain an upper bound of the minimum volume. First, the maximum of the products can be split between product of the maxima; second, the indicator function can be dropped. As a consequence, the optimization variable $t$ can be dropped too, since it only affects the objects' relative velocity and the binning indicator. Third, the maximum impact velocity can be approximated by doubling the circular velocity at the lowest altitude, $h_{min}$, considered by the software. All the operations individually are conservative, leading to the following upper bound:
\begin{equation}
\label{eq:unbiasedCube_conservative}
    dV_{min} \geq {2\sqrt{\frac{\mu}{R_e+h_{min}}}\,\max_{i,j}\sigma_{i,j}}\, \Delta t\,,
\end{equation}
where $\mu$ is the gravitational parameter of Earth and $R_e$ is its equatorial radius.
\begin{figure}
    \centering
    \includegraphics[width=0.5\linewidth]{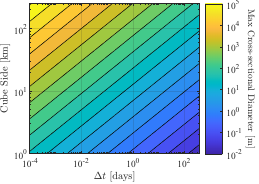}
    \caption{Maximum cross-sectional diameter to avoid bias as a function of time-step and cube side.}
    \label{fig:diameter_vs_dt_and_cube}
\end{figure}
Figure ~\ref{fig:diameter_vs_dt_and_cube} shows the maximum cross-sectional diameter allowed to avoid bias as a function of a cube side and integration time-step. The figure is obtained rearranging \eqref{eq:unbiasedCube_conservative} and setting $h_{min}=0$. Similarly, Fig.~\ref{fig:dt_and_cube} shows the valid set, in green, of time step and cube side, for a fixed value of the cross-sectional diameter set to 50~m.
On the other hand, one does not want too large a cube because it would dilute the probability of collision. Assume, for example, that the cube side were twice the diameter of Earth. Naturally, most of the involved volume would be empty all the time, since the cube would include regions like the interior of the Earth. Hence, the probability of collisions would always be divided by an overly conservative volume, introducing, also in this case, a bias. To mitigate bias caused by too large a cube, an empirical study beyond the scope of this research is required, and the interested reader is referred to Lewis et al.~\cite{LewisCUBElimitations2019} and to Facchinetti~\cite{facchinetti2023analysis}.
\section{Rare event simulation}
Rare event simulation~(RES) is a class of variance reduction methods for sampling-based estimators. This section reports the theoretical background necessary for this research, but is in no way exhaustive, and is only limited to importance sampling. The interested reader may refer to L'Ecuyer et al.~\cite{lecuyer2009splitting} and to Beck and Zuev~\cite{beck2015rare} for more details. The general problem of RES is to estimate the probability of an event that may otherwise rarely happen in simulation:
\begin{equation}
    p_E=P(U\in E) = \int_E p(u)du,
\end{equation}
where $p_E$ is small.
The rare event probability can be computed by standard MC simulations:
\begin{equation}
    \hat{p}_E = \frac{1}{N}\sum_{n=1}^N I^E(n)
\end{equation}
where $N$ is the number of samples used by the MC simulation, and $I^E(n)$ is 1 if event $E$ happened in simulation $n$, and zero otherwise.
Whether an event happens or not can be modeled by a Bernoulli distribution; therefore, the variance of $\hat{p}_E$, the estimate of $p_E$, is
\begin{equation}
    \text{Var}\left[\hat{p}_E\right] = \frac{1}{N}p_E\left(1-p_E\right).
\end{equation}
The standard deviation $\sigma_{\hat{p}_E}$ of the estimate, relative to the probability $p_E$ itself is thus
\begin{equation}
    \frac{\sigma_{\hat{p}_E}}{p_E} = \frac{1}{\sqrt{N}}\,\sqrt{\frac{\left(1-p_E\right)}{p_E}}\approx \sqrt{\frac{1}{N\,p_E}},
\end{equation}
where the approximation is valid since we assume $p_E$ small. Thus, evaluating the probability of a rare event with high relative accuracy can be prohibitive for small $p_E$.

A common solution to this issue consists of twisting the known sampling probability distribution and then compensating accordingly via importance weights. The process is commonly known as importance sampling, or exponential twisting in some cases. The twisting should be such that the samples belonging to the event under investigation are included more frequently.
In general, one samples from a twisted importance sampling distribution $q(u)$:
\begin{equation}
p_E = P(U\in E) = \int_E q(u)\frac{p(u)}{q(u)}du = \int_E p(u)du,
\end{equation}
where the weighing function ${p(u)}\,{q^{-1}(u)}$ ensures that the estimate remains unbiased. The estimate can be computed by MC, sampling from the twisted distribution instead of the original distribution:
\begin{equation}
    \hat{p}_E^q = \frac{1}{N}\sum_{n=1}^N I^E(n) \left( \frac{p(u_n)}{q(u_n)}\right),
\end{equation}
where the superscript $(\cdot)^q$ indicates that the estimate has been obtained sampling from $q(u)$. Note that, to avoid bias, $q(u)$ must be nonzero for every $u$ such that $p(u)\,I(u)>0$.

The variance of the estimate is thus now a function of $q(u)$, and it can be arbitrarily reduced, depending on how closely $q(u)$ resembles the set $E$:
\begin{equation}
    \text{Var}\left[\hat{p}^q_E\right] = \frac{1}{N} \left(\mathbb{E}_q\left[I^E(u) \frac{p^2(u)}{q^2(u)}\right]-p_E^2\right).
\end{equation}
Ideally, if the set $E$ were known, one could choose the optimal importance sampling distribution $q(u)$ such that the variance of the estimate would be zero:
\begin{equation}
    q_{opt}(u) = p(u|E) = \frac{ I^E(u) p(u)}{p_E}.
\end{equation}
Naturally, if $I^E(u)$ and $p_E$ were known, one would not need to estimate $p_E$.

\section{Rare Event Simulation for Collisions}
Recall that in MOCAT-MC the occurrence of a collision is modeled as a Bernoulli distribution with probability $p_{c,ij}$ computed according to Eq.~\eqref{eq:pc_cube}.
Here, the goal is no longer that of estimating a probability, but various metrics of the simulation, such as number of collisions, number of objects in orbit, and distribution of those objects. We focus specifically on the number of objects, since that is the major factor contributing to the number of future collisions.

\begin{figure}
    \centering
    \includegraphics{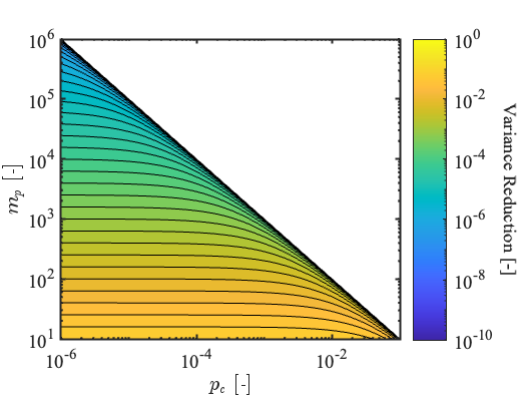}
    \caption{Ratio between modified variance and original variance in the number of objects generated by a single collision.}
    \label{fig:v_r_m_p_p_c}
\end{figure}

Let us assume for now that the simulation involves a single collision. One can multiply $p_{c,ij}$ by $m_p$, a collision probability multiplier.
To compensate, if the collision happens, the overall simulation would need to be weighed by the importance sampling factor $m_p^{-1}$.
Therefore, the expected number of simulated fragments $n_{f,s}$ and its variance are
\begin{align}
\label{eq:expvalue}
    \mathbb{E}_{m_p}& \left[n_{f,s} \right] = m_p\,p_{c,ij}\frac{n_{f,g}}{m_p};\\
\label{eq:vr}   \text{Var}_{m_p}&\left[n_{f,s} \right] = (1-m_p p_{c,ij})\,m_p p_{c,ij}\left(\frac{n_d}{m_p}\right)^2;
\end{align}
where $n_{f,g}$ is the number of fragments generated by the NASA SBM, and the subscript $(\cdot)_{m_p}$ implies that the statistics are computed sampling from a Bernoulli distribution whose parameter is multiplied by $m_p$. Note that the expected number of fragments $n_{f,s}$ is unaffected by the operation, since $m_p$ simplifies in Eq.~\eqref{eq:expvalue}. For small values of $p_{c,ij}$, the variance in the number of objects is reduced by $m_p$.
Figure~\ref{fig:v_r_m_p_p_c} shows the theoretical reduction in variance as a function of original probability of collision $p_c$ and multiplying factor $m_p$. Note that as long as $p_c \, m_p\ll 1$, the reduction in variance is approximately inversely proportional to $m_p$; however, once $p_c \, m_p$ approaches 1, which occurs when reaching the diagonal line splitting the figure, the overall variance tends to 0, dividing the original variance by a factor that tends to infinity.

\subsection{Multiple Collisions}
Let us consider now the case in which multiple collisions occur in a single simulation.
In such a case, in order to estimate the number of fragments generated, the weight $w_n$ to be applied to the $n^\text{th}$ simulation is
\begin{equation}
    w_n = \frac{1}{N}\,\prod_{n=1}^{N_{c,n}} I_c(i)\frac{1}{m_p},
\end{equation}
where $N_{c,n}$ is the number of collisions evaluated in simulation $n$, and $I_c(i)$ is 1 if the $i^{\text{th}}$ collision has occurred, and is 0 otherwise. Conversely, for traditional MC, all weights $w_n$ would be equal to $N^{-1}$. 
The weight $w_n$ is exponentially decreasing with the number of occurred collisions, which in turn increases for large $m_p$ and for long duration simulations. Note that there is a feedback loop that complicates the actual proper probability distribution: the twisting of probabilities generates more fragments, which then generate more collisions.
As an additional problem, since many more collisions occur, the number of generated objects can become excessively large in every simulation, increasing computational and memory requirements and leading to less effective outcomes than expected.

\subsection{Fragments Sampling}
We solve the above-mentioned problems in two steps: 1) instead of affecting the weights of the MC runs, we augment the debris fragments with a probability of existence equal to $m_p^{-1}$, and 2) instead of propagating all debris, we randomly sample from them. Step 1) of the solution solves the problem of the interference between additional fragments and new collisions, as well as the issue of exponentially diminishing weights. Step 2) mitigates, or entirely neutralizes, the exponential growth in computational complexity.
The sampling of fragments can be tuned such that the probability of existence $p_{e,k}$ after the sampling is a design parameter $p_{e}$ if the fragment is sampled. Note that $1\geq p_e \geq m_p^{-1}$. Hence
\begin{equation}
    p_{e,r}=\begin{cases}
p_e, & \text{with probability }\frac{p_e}{m_p}\\
0, & \text{otherwise}
\end{cases}.
\end{equation}
If $p_{e,r}=0$, the corresponding satellite is removed from the simulation.

Sampling the fragments as a sequence of independent Bernoulli distributions leads to a binomial distribution, which also has non-zero variance. The variance of the number of fragments to be retained $n_{f,r}$ is
\begin{equation}
    \text{Var}\left[n_{f,r}\right] = n_{f,g}\left(1-\frac{p_e}{m_p}\right)\,\frac{p_e}{m_p}.
\end{equation}
It is therefore preferable to first compute the expected value of $n_{f,r}$ by modifying Eq.~\eqref{eq:expvalue} to account for the case where the used defines $p_e < 1$:
\begin{equation}
    n_{f,r} = \mathbb{E}_{m_p} \left[n_{f,s} \, p_e\right] = p_e\,p_{c,ij}{n_{f,g}}.
\end{equation}
Then, a random permutation of $n_{f,r}$ (randomly rounded to an integer) objects is sampled from the generated fragments. To reduce the variance of the mass of objects in orbit, the fragments are sorted by mass and are sampled at regular intervals in position, with integer rounding applied as necessary.

\subsection{Existence of Parents}
Insofar, the fate of the parent objects in a potential collision has not been considered. 
%Two approaches are proposed, which can provide different advantages depending on the simulation. In both cases, if the collision has not occurred despite the probability twisting, nothing happens.
%If the collision has occurred, one approach consists of setting the probability of persistence of the parents as $1/m_p$.
%Whilst this approach does not generate a large variance in the number of objects, since the parents are only two per collision, it maintains a large variance in the mass in orbit. That is because the parents are generally large, and always larger than all the fragments they generate (because of the variance reduction).
In this work, we augment the parents with a probability of existence too. However, given their large mass, and limited number, we choose to not randomly sample their existence: their probability of existence is just decreasing with every collision. Thus:
\begin{equation}
    p_{e,i}^+ = p_{e,i}^- \left(1-m_p^{-1}\right),
\end{equation}
where the superscripts $(\cdot)^-$ and $(\cdot)^+$ indicate that the probability of existence refers to before and after the collision, respectively. Note that the update in $p_e$ only occurs if the collision has been simulated. The probability of existence of the parents multiplies all future probabilities of collision in which they are involved.

While this approach reduces the variance in the total mass in orbit, it also leads to maintaining objects forever until they deorbit. Thus, for large simulations with many objects, and where many collisions happen, the latter approach might result in computational inefficiencies, and one may want to sample the existence of the parents after their probability of existence goes below a predefined threshold.

\section{Quasi Zero-Variance Collisions}
Up until now we have only considered the case where $m_p$ is a factor constant throughout the simulation, independent of the collision under considered. However, as shown in Fig.~\ref{fig:v_r_m_p_p_c}, the optimal value for $m_p$ is strongly dependent on the collision probability $p_c$. The optimal value can be derived from Eq.~\eqref{eq:vr}:
\begin{equation}
 m_{p,ij}^{*} = p_{c,ij}^{-1} ,  
\end{equation}
where the superscript $(\cdot)^{*}$ stands for optimality.
In the ideal case, the variance of the collision event is zero. That, in turn, would lead to zero variance in the number of generated objects, except for integer rounding. Note that while the number of collision fragments is quasi-deterministic, the overall outcome remains stochastic, since the fragments have to be sampled from the output of the SBM, and have random mass, surface area, and velocity. However, a reduction in the variance of the number of objects is a key factor in reducing the overall variance of the system.

\subsection{Compute Time Management}
The quasi zero-variance solution implies that all potential collisions need to be generated with the SBM, regardless of how small the corresponding probability of collision is. Generating all fragments from all possible collisions may then become the bottleneck of the computation.
We find that a threshold on $m_{p,ij}$ is effective in decreasing the computational burden.
Specifically, it is helpful to have $m_{p,ij}$ increase with the total mass involved in the collision, so that the more massive collisions,
which would generate large numbers of fragments,
are likely to be computed,
whereas small collisions are likely to be skipped. The filter used is quadratic in the mass:
\begin{equation}
    m_{p,ij} = \min \left(\frac{1}{p_{c,ij}}, \frac{1}{k_f} (m_i+m_j)^2\right),
\end{equation}
where $k_f$ is a tuning parameter.
Different formulations of the filter may be useful, but we find that the proposed equation greatly reduces the computational burden while maintaining the variance reduction. %Note that while additional factors may influence the number of fragments generated, such as the collision velocity, they all already affect the probability of collision $p_{c,ij}$ according to Eq.~\eqref{eq:pc_cube}; thus, their importance is already weighed in the probability of collision.

\subsection{Variance of Post Mission Disposal Maneuvers}
The PMD is another major source of variance, albeit orders of magnitude smaller than the collisions. Similarly to what is done with the sampling of the fragments, the variance of PMD maneuvers can be reduced by counting the expected number of successful maneuvers between two consecutive failures as the inverse of the failure probability of the PMD. Failures are then set to always and only occur after the expected number of successful PMDs. Integer correction is accounted for if needed.
This approach reduces the variance in the number of derelict spacecraft in orbit from $N_{PMD}\, P_{PMD}(1-P_{PMD})$, where $N_{PMD}$ is the number of PMD maneuvers attempted, to less than 1.

\section{Standard Deviations Thresholds}

The approaches for variance reduction that have been proposed so far do not provide any information on the standard deviation of the actual distribution of the simulations. In many cases, one may be interested in the worst case scenarios, or in certain percentiles of realizations, \textit{e.g.}, the Value at Risk~(VaR).
For example, one might be interested in the outcome of a chosen policy under the 3$\sigma$ pessimistic bounds, as a practical approximation of the worst-case scenario.
If one wants to evaluate the 3$\sigma$ worst case threshold, or 3$\sigma$ VaR, the standard deviations extracted from the distribution of MOCAT-QD would not apply, since its standard deviation is artificially reduced. Hence, this section introduces MOCAT-QD-VaR, a version of MOCAT-QD that can compute the VaR for $n_\sigma$ standard deviations \textit{with a single simulation}, where $n_\sigma$ is a user-defined parameter.  
In MOCAT-QD-VaR we only look at the major source of uncertainty, the collisions. This design choice is made for two reasons: 1) the total number of fragments generated from a collision is the largest source of uncertainty in the total number of objects in space, and 2) affecting more than one source of uncertainty by a factor of $n\,\sigma$ standard deviations would lead to an overall deviation of the simulation from the mean that is larger than $n\,\sigma$. 

Let us start by assuming that the entire system consists of a single collision. In that case, one can simply artificially increase the probability of collision $p_{c,ij}$ by $n_\sigma$ standard deviations. For a single Bernoulli distribution:
\begin{equation}
\label{eq:p_t_single}
    p_t = p_c + n_\sigma\sqrt{(1-p_c)p_c},
\end{equation}
where $p_t$ stands for the twisted probability of collision.
Note that the formula only makes sense when $p_c$ is small, which is generally the case when using the cube method. Otherwise, the resulting probability of collision would be larger than 1, causing a bias in the simulation.
When looking at the combination of independent Bernoulli distributions, the twisting probability depends on the number of total sampling events. Assuming, for now, same $p_c$ for all events:
\begin{equation}
\label{eq:p_t_many}
    p_{t,avg} = p_c + n_\sigma\sqrt{\frac{(1-p_c)p_c}{N_c}},
\end{equation}
where $p_{t,avg}$ is an average of $p_t$ over the whole simulation.
If, instead, the simulation consists of multiple collisions, all with different values of $p_c$:
\begin{equation}
\label{eq:p_t_many_varying}
    p_{t,avg} = p_c + n_\sigma\sqrt{\frac{\sum_{k=1}^{N_c}(1-p_{c,k})p_{c,k}}{N_c^2}}.
\end{equation}
The above equation does not account, however, for the fact that different collisions generate different numbers of fragments, depending on the objects involved.
In reality, we are interested in the standard deviation of the total number of fragments generated in a simulation.
Let $\sigma_{n_f,1:l}$ be the cumulative standard deviation of fragments generated by all conjunctions from the first to the $l$\textsuperscript{th}: 
%\begin{equation}
%\label{eq:p_t_many_varying_and_frags}
%    p_t = p_c + n_\sigma\sqrt{\frac{\sum_{k=1}^{N_c}(1-p_{c,k})p_{c,k}n^2_{f,g,k}}{N_c^2}}.
%\end{equation}
\begin{equation}
\label{eq:sigma_frags}
    \sigma_{n_f,1:l} = \sqrt{{\sum_{k=1}^{l}(1-p_{c,k})\,p_{c,k}\,n^2_{f,g,k}}},
\end{equation}
where $n_{f,g,k}$ is the number of fragments generated by the SBM at the conjunction $k$.
As a consequence, the average number of fragments to be added at conjunction $l$ to keep 1 standard deviations from the mean is
\begin{equation}
\label{eq:sigma_nfl}
    \sigma_{n_f,l} = \sqrt{{\sum_{k=1}^{l}(1-p_{c,k})\,p_{c,k}\,n^2_{f,g,k}}}-\sqrt{{\sum_{k=1}^{l-1}(1-p_{c,k})\,p_{c,k}\,n^2_{f,g,k}}}.
\end{equation}
The corresponding twisted probability for collision $l$ is then
\begin{equation}
    p_{t,l} = p_{c,l} + n_\sigma\,\frac{\sigma_{n_f,l}}{n_{f,l}},
\end{equation}
where the division by $n_{f,l}$ is
so that the expected number of fragments remains $p_{c,l}\,n_{f,g,k} + n_\sigma\,\sigma_{n_f,l}$.
To be able to compute $\sigma_{n_f,l}$ one needs to compute $n_{f,g,k}$ for $k=1, ..., l$ , which can be prohibitive for large simulations. An alternative to computing all those collisions is to estimate the variance of all collisions just from actually occurred collisions, since, by definition, the sample variance tends to the actual variance:
\begin{equation}
\label{eq:var_bernoulli}
\lim_{l\to\infty}\frac{{\sum_{k=1}^{l-1}(1-p_{c,k})\,p_{c,k}\,n^2_{f,g,k}}}{l-1} = \frac{\sum_{k=1}^{l-1}\left(I(k)n_{f,g,k}-\frac{\sum_{k=1}^{l-1}I(k)n_{f,g,k}}{l-1}\right)^2}{l-1}
\end{equation}
where $I(k)$ is 1 if collision $k$ has occurred, and zero otherwise. Note that since $n_{f,g,k}$ is always multiplied by $I(k)$, it is not necessary to compute it when the collision has not been simulated, saving computational time. Equation~\eqref{eq:var_bernoulli} can be used to approximate the rightmost addend in Eq.~\eqref{eq:sigma_nfl}. Computing Eq.~\eqref{eq:sigma_nfl} still requires knowledge of $n_{f,l}$, which are the number of fragments that would be generated by the collision currently being sampled. Having to compute that would amount to having to compute all collisions of the simulation. However, computing it can be avoided by a simple trick: instead of actually augmenting the probability of collision, we can augment the number of generated fragments if the collision has occurred. This allows to obtain an $n_\sigma$ deviations scenario without having to compute all fragments from all potential conjunctions.

\section{Simulation Results}
We provide several results that validate the claim that the proposed method reduces the variance of all-vs-all simulators while introducing bias that is negligible for most purposes. For now, results only focus on the total number objects in space.
First, we evaluate the ``linear" region of $m_p$ according to Fig.~\ref{fig:v_r_m_p_p_c}. In such a region, we expect that a constant value of $m_p$ will lead to an equivalent reduction in variance. Then, we look at the quasi-deterministic approach, MOCAT-QD, and determine how much the variance can be reduced as a function of simulation parameters. Finally, we validate MOCAT-QD-VaR.
For all the following simulations a cube of 50~km per side is selected, coupled with a time-step of 5 days.

\subsection{Constant Factor Reduction}
First we analyze the results for the region where $m_p$ is a constant, relatively small, value.
Note that to keep the estimate unbiased, one needs that the product $m_p \, P_{c,ij}$ is capped at 1. The analysis focuses on the average total number of objects. The ground truth is generated with an estimate obtained by averaging 500 runs of MOCAT-MC.
\begin{figure}
    \centering
    \includegraphics{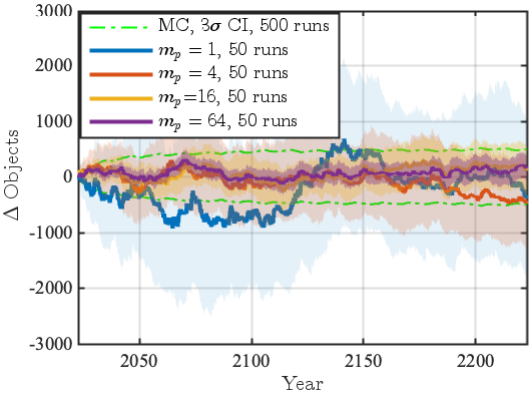}
    \caption{Difference in average number of objects between MOCAT-MC with different values of $m_p$ (50 runs), and standard MOCAT-MC with $m_p=1$ (500 runs).}
    \label{fig:var_red1_4_16_64}
\end{figure}
Fig.~\ref{fig:var_red1_4_16_64} shows the deviations of the new approach from the MOCAT-MC with 500 runs. The green lines delimit the $3\sigma$ confidence interval~(CI) of the estimate of the mean obtained with the 500 MOCAT-MC runs. The figure also includes an estimate obtained from only 50 runs of MOCAT-MC with $m_p=1$ as a comparison. Interestingly, the latter is the only one whose average ever goes outside of the 3$\sigma$ bounds obtained with 500 runs. Just by eye inspection, it is possible to see that the CI for $m_p=1$ is approximately twice as large as the CI for $m_p=4$, which themselves are approximately twice, and four times, larger than the intervals for $m_p=16$ and $m_p=64$, respectively. Towards the end, the mean estimate for the standard MOCAT-MC, and MOCAT-QD with $m_p=4$, almost drifts outside the 3$\sigma$ CI of the estimate obtained with 500 runs. On the other hand, the two estimates with larger variance reduction remain very close to being unbiased.
\begin{figure}
    \centering
    \includegraphics{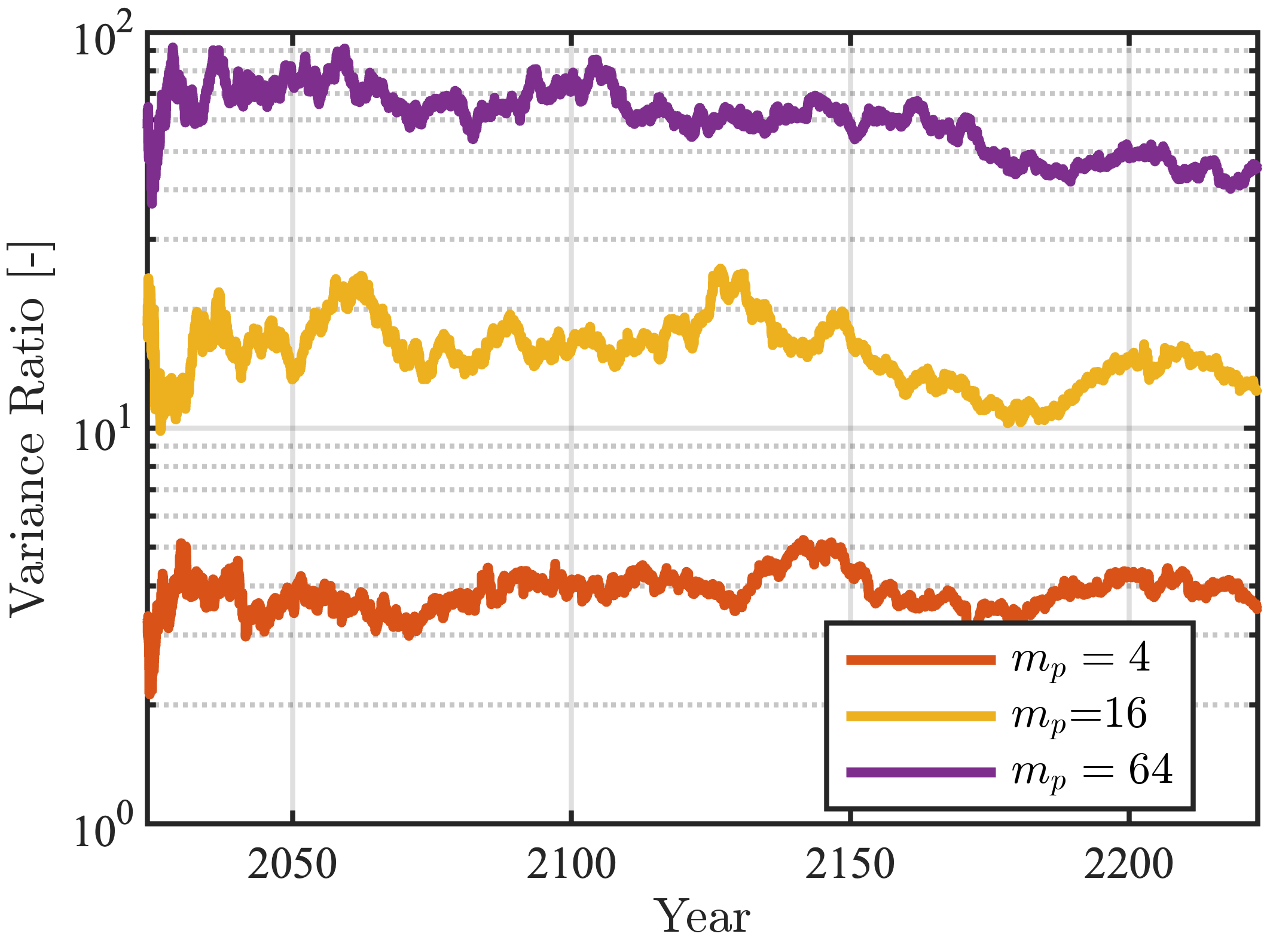}
    \caption{Variance reduction factor for MOCAT-MC with different values of $m_p$, averaged over 50~runs.}
    \label{fig:var_red1_4_16_64_var}
\end{figure}
Fig.~\ref{fig:var_red1_4_16_64_var} shows the actual variance reduction for different values of $m_p$. On average, the reduction factor is between 4\% and 8\% less than the preset value of $m_p$, with less relative efficiency for larger $m_p$. This effect is likely due to secondary sources of uncertainty, such as random initial conditions, whose variance is not reduced in the simulation of this section. Overall, the empirical results validate the theoretical results from Fig.~\ref{fig:v_r_m_p_p_c}, albeit with slightly less efficiency than expected.
As a final remark, note that the variance reduction leads to $m_p$ times more collisions being computed. This in turn can increase computational cost, but for these relatively small values no relevant time increase is noticed: even with $m_p=64$, the computational bottleneck is still the orbit propagation. This is no longer the case with MOCAT-QD, if no filter is applied.

\subsection{Quasi-Deterministic MOCAT}

\subsubsection{Random Initial Conditions.}
The first scenario includes random initial conditions.
\begin{figure}
    \centering
    \includegraphics{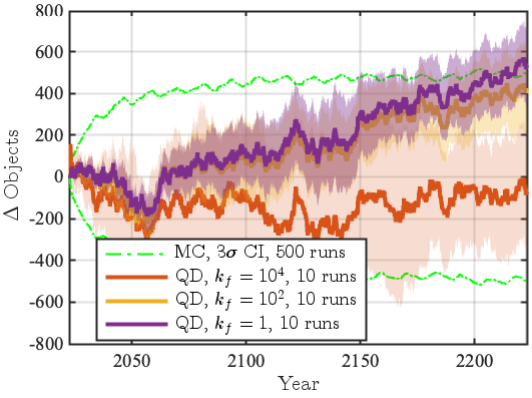}
    \caption{Difference in average number of objects between different instances of MOCAT-QD and MOCAT-MC (500 runs), and corresponding $3\sigma$ CI.}
    \label{fig:errors_5dt_50km}
\end{figure}
Figure~\ref{fig:errors_5dt_50km} shows the evolution in time of the difference between MOCAT-MC, averaged over 500 runs, and different instances of MOCAT-QD, averaged over only 10 runs. All runs include 3$\sigma$ standard deviations for estimates of the mean. The 3$\sigma$ uncertainty for MOCAT-MC is in green and centered over 0. At all times, the runs with MOCAT-QD and MOCAT-MC stay within each others' 3$\sigma$ uncertainties. However, for MOCAT-QD versions that have lower values of $k_f$, and hence where the variance reduction is more aggressive, their mean deviates from that of MOCAT-MC, at a quite steady pace. Interestingly, this trend is the opposite of the one noticed in Fig.~\ref{fig:var_red1_4_16_64}, where instead the simulations leading to less variance reduction were more likely to introduce a bias.
\begin{figure}
    \centering
    \includegraphics{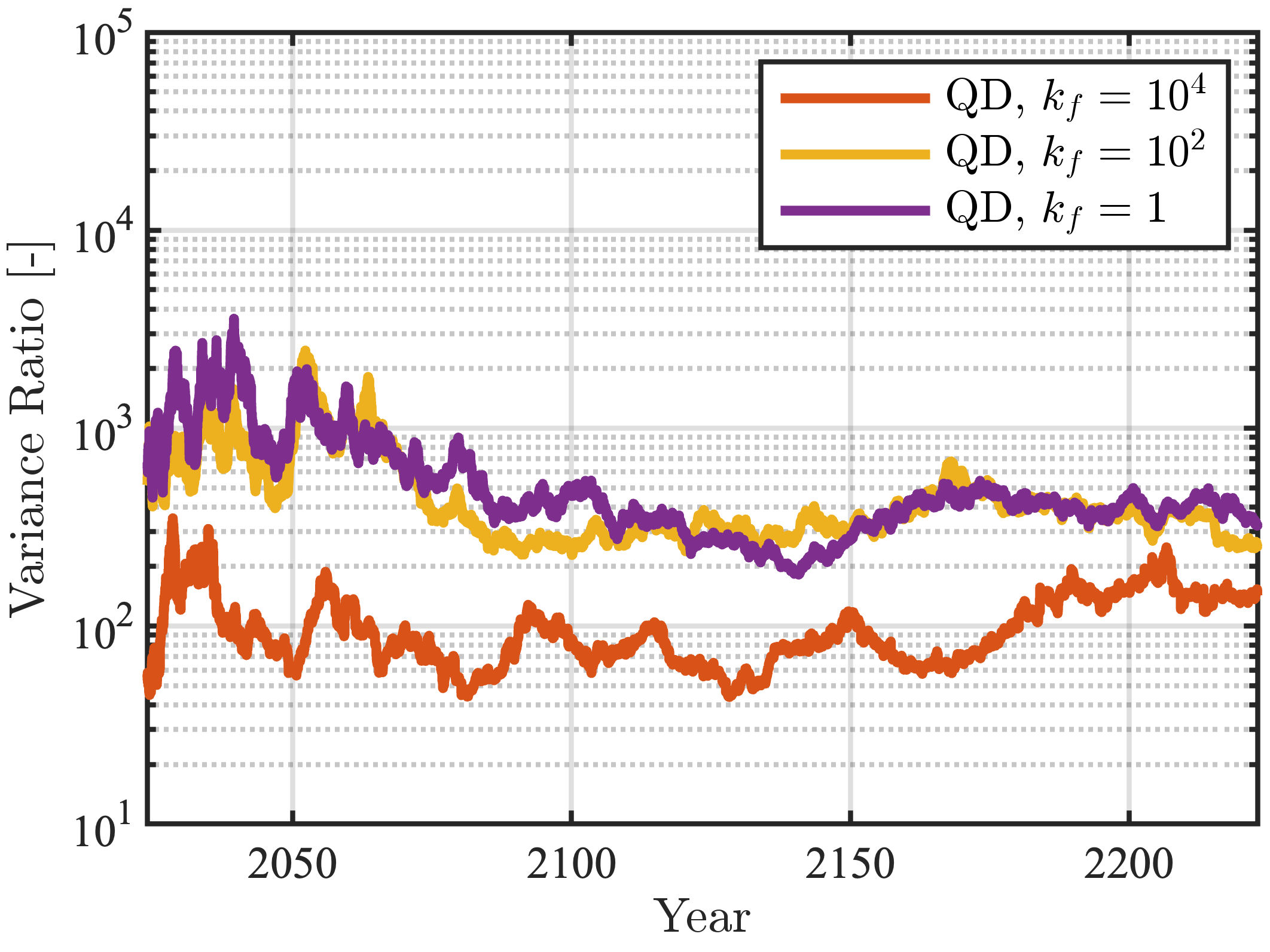}
    \caption{Variance reduction factor between different instances of MOCAT-QD and MOCAT-MC.}
    \label{fig:var_red_Inf_5dt_50km}
\end{figure}
Fig~.\ref{fig:var_red_Inf_5dt_50km} shows the evolution of the effective variance reduction in time. As expected, the largest value of $k_f$ provides the smallest variance reduction. Neglecting the first year, where the estimates are very noisy, the variance is reduced by a factor between 44 and 350, with an average over time of 84, and with a final value of 140. For the other two values of $k_f$, which can be discussed together since their behavior is similar to one another, the variance is reduced by a factor ranging between 200 and 3,000; on average, a choice of $k_f=10^2$ and $k_f=1$ lead to, respectively, a reduction factor of 387 and 425. These results empirically validate the proposed approach. For the majority of simulation time, the average from just two runs of MOCAT-QD with $k_f = 100$ is more precise than the average of 500 MOCAT-MC runs. At many epochs, even a single run of MOCAT-QD is more precise.

\subsubsection{Averaged Initial Conditions.}
\begin{figure}
    \centering
    \includegraphics{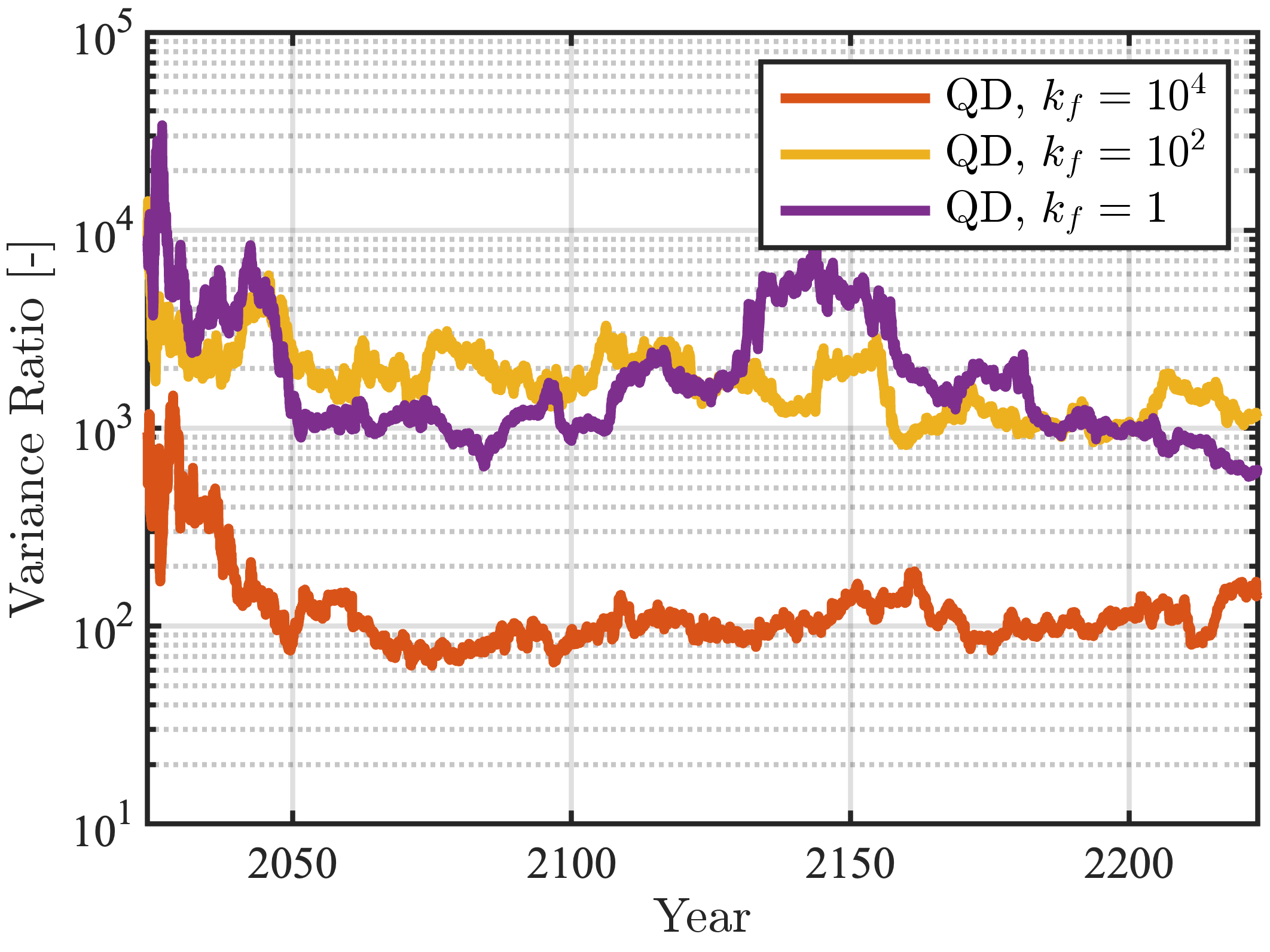}
    \caption{Variance reduction factor between different instances of MOCAT-QD and MOCAT-MC. Initial conditions for MOCAT-QD are deterministic.}
    \label{fig:var_red_Inf_5dt_50km_detIC}
\end{figure}

This subsection empirically demonstrates that by using averaged (and thus deterministic) initial conditions, one does not introduce any noticeable bias, and greatly reduces the variance further. %Hence, we now look at a case in which the initial conditions are deterministic, to determine an upper bound of how well the method might perform in case variance reduction techniques for initial conditions were included.
\begin{figure}
    \centering
    \includegraphics{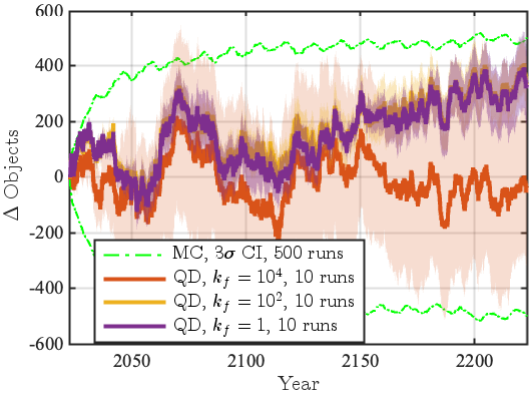}
    \caption{Difference in mean number of objects between different instances of MOCAT-QD and standard MOCAT-MC (500 runs), and corresponding $3\sigma$ CI. Initial conditions for MOCAT-QD are deterministic.}
    \label{fig:var_red_inf_detIC}
\end{figure}
Figure~\ref{fig:var_red_inf_detIC} shows that using deterministic conditions does not introduce any additional noticeable bias, but might instead reduce it. A possible explanation for this is that choosing the average every time might be a more representative value than 10 random samples.
Figure~\ref{fig:var_red_Inf_5dt_50km_detIC} shows that the variance reduction is now greatly increased for some values of $k_f$, substantiating the idea that the randomness in initial conditions was indeed a limiting factor for the previous results. Once again the best trade-off is obtained with $k_f=100$. In this case, except for a short period, the variance is always reduced by a factor larger than 1,000, averaging 1,594. On the other hand, deterministic initial conditions only marginally affect MOCAT-QD with $k_f=10,000$, leading to an average variance reduction of 108: the implication is that in that case the largest source of uncertainty remains the collisions.
It is worth noting that while setting deterministic initial conditions considerably reduces, in relative terms, the variance of MOCAT-QD, that is not the case for MOCAT-MC, where the variance reduction is of only 0.6\%. That is because in MOCAT-QD with small $k_f$ the variance bottleneck is the randomness of initial conditions, whereas in MOCAT-MC the variance bottleneck is the randomness of collisions.

Table~\ref{tab:compute_time} shows the computational time required by the different approaches and their corresponding reductions in variance. As MOCAT-QD with $k_f=10,000$ requires approximately the same compute as MOCAT-MC, but produces a variance that is 108 times smaller, it is about 100 times more efficient than MOCAT-MC. MOCAT-QD with $k_f=100$ is likely the overall most efficient version, improving on the efficiency of MOCAT-MC by a factor of around 1,090 times when taking into account the increased compute time.

\begin{table}[]
    \centering
    \begin{tabular}{c|c|c|c|c}
        MOCAT Version & $k_f$ & Compute time per run [s] & Var. red. & Var. red., det. IC\\
        \hline
       MC  &  N/A & 252.96 & 1 & 1.0055\\
       QD  &  1 & 773.64 & 424.53& 1,420.5\\
       QD  &  100 & 369.52 & 387.41& 1,594.5\\
       QD  &  10,000 & 272.89 & 83.73& 108.17 \\
    \end{tabular}
    \caption{Computation time, averaged over multiple runs, and variance reduction, averaged over multiple runs and over all epochs.}
    \label{tab:compute_time}
\end{table}

\subsection{Value at Risk MOCAT-QD}
\begin{figure}
    \centering
    \includegraphics[width=0.5\textwidth]{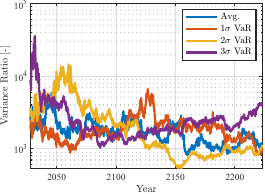}
    \caption{Variance reduction obtained with MOCAT-QD-VaR, $k_f=10^2$.}  \label{fig:var_red_1vsInfQD1sigms_2dt_30km_50_runs_detIC_var}
\end{figure}

This section provides results obtained with the formulation of MOCAT-QD that is capable of approximating the VaR with a single run.
\begin{figure}
    \centering
    \includegraphics{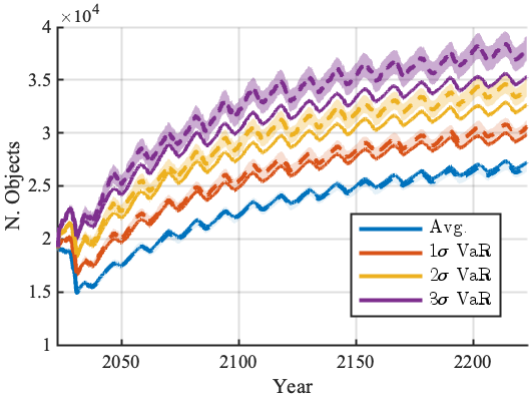}
    \caption{Mean and 3$\sigma$ CI of the VaR computed from 500 runs of MOCAT-MC (dashed lines), and from 10 runs MOCAT-QD-Var, $k_f=10^2$ (continuous lines).}  \label{fig:var_red_1vsInfQD1sigms_2dt_30km_50_runs_detIC}
\end{figure}
Figures~\ref{fig:var_red_1vsInfQD1sigms_2dt_30km_50_runs_detIC} and ~\ref{fig:var_red_1vsInfQD1sigms_2dt_30km_50_runs_detIC_var} show the results obtained for 1, 2, and 3 standard deviations. The plot shows the mean of the VaRs obtained with MOCAT-MC, and the corresponding results obtained with MOCAT-QD. MOCAT-MC is estimated over 500 runs, whereas MOCAT-QD is evaluated over 10 runs. Note that the figure includes the CI from MOCAT-QD-Var, but they are so small that they can barely be seen. MOCAT-QD-Var is optimistic, and its estimate lies outside of the 3$\sigma$ CI obtained with 500 runs of MOCAT-MC. For the 3$\sigma$ VaR, the bias at the end of the simulation is such that little more than 90 MOCAT-MC runs are sufficient to provide a more accurate estimate than MOCAT-QD-VaR with a confidence of 99.7\%. For the 2$\sigma$ Var, approximately 125 MOCAT-MC runs are needed instead, and for the 1$\sigma$ VaR, more than 500. In both cases, the bias begins inside the 3$\sigma$ CI of the estimates from MOCAT-MC, and then grows outside. Hence, for short time predictions, the bias is even less relevant.
The variance reduction factor for the $3\sigma$ VaR at is 4,300 at the end of the simulation. The reduction is much larger than when estimating the mean because the estimate of the standard deviation is itself noisy, and when it is multiplied by a factor of 3, the variance is greatly amplified. On the other hand, with MOCAT-QD-VaR, no multiplications of noisy estimates are needed. Hence, MOCAT-QD-VaR is the software most successful in reducing the variance; nonetheless, it also introduces a visible bias. We believe that the bias is likely caused by one, or both, of the following factors: 1) the PMD is not included in the probability twisting, and 2) the distribution of the total number of objects is non-Gaussian, and is actually quite heavy-tailed, because of the feedback loop where past collisions increase the probability of future collisions. In other words, the collisions are not independent events.

\section{Conclusions}
This paper contributes to the field of orbital evolutionary models in three ways.
First, it provides a theoretical justification for the choice of time-step and cube size of MOCAT-MC, which is a critical tuning problem that has been empirically studied in literature, but for which, to the best of the authors' knowledge, no theoretical explanation has been provided. While the tuning is not validated here, the theoretical results qualitatively match empirical literature.
Second, the principal contribution of this paper is the introduction of MOCAT-QD, an all-vs-all simulator that provides high-fidelity, low variance solutions to the problem of simulating the long-term environment of LEO. The variance is reduced by a factor between 200 and 1,500 for the scenarios investigated, with only a 50\% increase in computation time. The introduced bias, if any, is so small that it is not visible even when comparing to the average from 500 runs of MOCAT-MC. The variance reduction is in some cases more significant if the simulations are for a short amount of time. Such a reduction in variance allows for precise and efficient sensitivity analyses, mission sustainability assessments, and risk evaluations.
Third, this paper also introduces a version of MOCAT-QD that provides a prespecified number of standard deviations from the mean. Hence, a 1, 2, or 3 $\sigma$ Value at Risk can be estimated with a single run.
Because the uncertainty on the standard deviation is multiplied by $n_\sigma$ when using MOCAT-MC, MOCAT-QD-VaR can be 3 times more efficient than the standard version of MOCAT-QD when seeking the $3\sigma$ VaR, leading to variance reduction factors of up to 4,300. However, likely because of the heavy-tailed distribution of all-vs-all simulators, MOCAT-QD-Var leads to biased results for the 2 and 3$\sigma$ VaR; the corresponding biases are such that, respectively, 125 and 90 MOCAT-MC runs would provide a more accurate estimate with 99.7\% confidence. Future work will focus on correcting this bias as well as on extending MOCAT-QD capabilities to include launches and rocket body explosions.
MOCAT-QD-Var is a key enabler for the evaluation of the VaR when performing policy evaluations and risk assessments.

\section{Acknowledgments}
Research was sponsored by the Department of the Air Force Artificial Intelligence Accelerator and was accomplished under Cooperative Agreement Number FA8750-19-2-1000. The views and conclusions contained in this document are those of the authors and should not be interpreted as representing the official policies, either expressed or implied, of the Department of the Air Force or the U.S. Government. The U.S. Government is authorized to reproduce and distribute reprints for Government purposes notwithstanding any copyright notation herein.

\bibliographystyle{AAS_publication}   % Number the references.

\begin{thebibliography}{10}

\bibitem{Kessler1978CollisionBelt}
D.~J. Kessler and B.~G. Cour-Palais, ``{Collision frequency of artificial satellites The creation of a debris belt},''  {\em Journal of Geophysical Research}, Vol.~83, No.~A6, 1978.

\bibitem{lifson2024low}
M.~Lifson, D.~Arnas, M.~Avenda{\~n}o, and R.~Linares, ``Low-Earth-Orbit Packing: Implications for Orbit Design and Policy,''  {\em Journal of Spacecraft and Rockets}, 2024, pp.~1--14.

\bibitem{Lucken2019CollisionDesign}
R.~Lucken and D.~Giolito, ``{Collision risk prediction for constellation design},''  {\em Acta Astronautica}, Vol.~161, Aug. 2019, pp.~492--501, 10.1016/J.ACTAASTRO.2019.04.003.

\bibitem{MITRI}
S.~Servadio, N.~Simha, D.~Gusmini, D.~Jang, T.~St.~Francis, A.~D’Ambrosio, G.~Lavezzi, and R.~Linares, ``Risk Index for the Optimal Ranking of Active Debris Removal Targets,''  {\em Journal of Spacecraft and Rockets}, Vol.~0, No.~0, 0, pp.~1--14, 10.2514/1.A35752.

\bibitem{nasaSTMevolve4}
N.~L. Johnson, P.~H. Krisko, J.~C. Liou, and P.~D. Anz-Meador, ``{NASA}'s New Breakup Model of EVOLVE 4.0,''  {\em Advances in Space Research}, Vol.~28, Jan. 2001, pp.~1377--1384, 10.1016/S0273-1177(01)00423-9.

\bibitem{legend2004}
J.~C. Liou, D.~T. Hall, P.~H. Krisko, and J.~N. Opiela, ``{LEGEND} – a Three-dimensional {LEO-to-GEO} Debris Evolutionary Model,''  {\em Advances in Space Research}, Vol.~34, Jan. 2004, pp.~981--986, 10.1016/J.ASR.2003.02.027.

\bibitem{Lewis2001DAMAGE:Framework}
H.~Lewis, ``{DAMAGE: a Dedicated GEO Debris Model Framework},''  Darmstadt, Germany, ESOC, Mar. 2001.

\bibitem{DELTA2}
C.~E. Martin, J.~E. Cheese, and H.~Klinkrad, ``{Space Debris Environment Analysis with DELTA 2.0},''  {\em International Astronautical Federation - 55th International Astronautical Congress}, 2004, 10.2514/6.IAC-04-IAA.5.12.5.04.

\bibitem{hanada2013orbital}
T.~Hanada, ``Orbital debris modeling and applications at Kyushu University,''  {\em Procedia Engineering}, Vol.~67, 2013, pp.~404--411.

\bibitem{ADEPT2020}
G.~Henning, M.~Sorge, G.~Peterson, A.~Jenkin, D.~Mains, and J.~McVey, ``Parameterizing large constellation post-mission disposal success to predict the impact to future space environment,''  {\em Journal of Space Safety Engineering}, Vol.~7, No.~3, 2020, pp.~171--177.
\newblock Space Debris: The State of Art, 10.1016/j.jsse.2020.07.025.

\bibitem{Jang2024Monte}
D.~Jang, D.~Gusmini, P.~M. Siew, A.~D'Ambrosio, S.~Servadio, P.~Machuca, and R.~Linares, ``A New Monte-Carlo Model for the Space Environment,''  2024.

\bibitem{medee}
J.-C. Dolado-Perez, R.~Di~Costanzo, and B.~Revelin, ``{Introducing {MEDEE} - a New Orbital Debris Evolutionary Model},''  {\em Proceedings of 6th European Conference on Space Debris}, Apr. 2013, pp.~22--25.

\bibitem{luca2}
J.~Radtke, S.~Mueller, V.~Schaus, and E.~Stoll, ``{LUCA2 - An enhanced long-term utility for collision analysis},''  {\em Proceedings of 7th European Conference on Space Debris}, 2017.

\bibitem{sharmaspace}
R.~K. Sharma and A.~A. Kumar, ``Space Debris and Space Situational Awareness Research Studies in ISRO,''  2024.

\bibitem{iadccomparison}
J.~Liou, A.~Anilkumar, B.~B. Virgili, T.~Hanada, H.~Krag, H.~Lewis, M.~Raj, M.~Rao, A.~Rossi, and R.~Sharma, ``Stability of the Future LEO Environment - an IADC Comparison Study,''  {\em Proceedings of 6th European Conference on Space Debris}, 2013, 10.13140/2.1.3595.6487.

\bibitem{LiouCUBE2003}
J.~C. Liou, D.~J. Kessler, M.~Matney, and G.~Stansbery, ``A New Approach To Evaluate Collision Probabilities Among Asteroids, Comets, And Kuiper Belt Objects,''  {\em Lunar and Planetary Science Conference}, 2003.

\bibitem{LiouCUBE2006}
J.~C. Liou, ``Collision activities in the future orbital debris environment,''  {\em Advances in Space Research}, Vol.~38, 2006, pp.~2102--2106, 10.1016/j.asr.2005.06.021.

\bibitem{LewisCUBElimitations2019}
H.~G. Lewis, S.~Diserens, T.~Maclay, and J.~P. Sheehan, ``Limitations of the cube method for assessing large constellations,''  {\em First International Orbital Debris Conference}, 2019.

\bibitem{facchinetti2023analysis}
G.~Facchinetti, ``Analysis and validation of the Cube Algorithm: assessing gas kinetic theory to model collision risk in long-term debris propagations,''  2023.

\bibitem{MOCAT-bin-AMOS2022}
D.~Jang, A.~D’Ambrosio, M.~Lifson, C.~Pasiecznik, and R.~Linares, ``Stability of the LEO Environment as a Dynamical System,''  {\em Advanced Maui Optical and Space Surveillance Technologies Conference}, 2022.

\bibitem{MOCAT-capacity-AMOS2022}
A.~D’Ambrosio, M.~Lifson, D.~Jang, C.~Pasiecznik, and R.~Linares, ``Projected Orbital Demand and LEO Environmental Capacity,''  2022.

\bibitem{mocat-ssem}
A.~D'Ambrosio, S.~Servadio, P.~M. Siew, and R.~Linares, ``Novel Source-Sink Model for Space Environment Evolution with Orbit Capacity Assessment,''  {\em Journal of Spacecraft and Rockets}, Vol.~60, 02 2023, 10.2514/1.A35579.

\bibitem{rodriguez2024towards}
V.~Rodriguez-Fernandez, S.~Sarangerel, P.~M. Siew, P.~Machuca, D.~Jang, and R.~Linares, ``Towards a Machine Learning-Based Approach to Predict Space Object Density Distributions,''  {\em AIAA SCITECH 2024 Forum}, 2024, p.~1673.

\bibitem{MOCAT3}
A.~D'Ambrosio, M.~Lifson, and R.~Linares, ``The Capacity of Low Earth Orbit Computed using Source-sink Modeling,''  {\em arXiv}, 2022, 10.48550/arXiv.2206.05345.

\bibitem{lecuyer2009splitting}
P.~L'Ecuyer, F.~Le~Gland, P.~Lezaud, and B.~Tuffin, {\em Rare Event Simulation using Monte Carlo Methods}, ch.~3, pp.~39--61.
\newblock John Wiley \& Sons, Ltd, 2009, 10.1002/9780470745403.ch3.

\bibitem{beck2015rare}
J.~L. Beck and K.~M. Zuev, ``Rare event simulation,''  {\em arXiv preprint arXiv:1508.05047}, 2015, 10.48550/arXiv.1508.05047.

\bibitem{zucchelli2024bayesian}
E.~M. Zucchelli and B.~A. Jones, ``A Bayesian Approach to Low-Thrust Maneuvering Spacecraft Tracking,''  {\em Journal of Guidance, Control, and Dynamics}, Vol.~1586--1601, Aug. 2024, 10.2514/1.G007849.

\bibitem{SZTAMFATERGARCIA2025}
Y.~Sztamfater-Garcia, M.~Sanjurjo-Rivo, G.~Escribano, H.~Molina-Bulla, and J.~Miguez, ``An approximate model for the computation of in-orbit collision probabilities using importance sampling,''  {\em Advances in Space Research}, 2025, https://doi.org/10.1016/j.asr.2024.12.074.

\bibitem{brouwer1959solution}
D.~Brouwer, ``Solution of the problem of artificial satellite theory without drag,''  {\em Astronomical Journal, Vol. 64, p. 378 (1959)}, Vol.~64, 1959, p.~378.

\bibitem{lyddane1963small}
R.~Lyddane, ``Small eccentricities or inclinations in the Brouwer theory of the artificial satellite,''  {\em Astronomical Journal, Vol. 68, p. 555 (1963)}, Vol.~68, 1963, p.~555.

\bibitem{Martinusi2015}
V.~Martinusi, L.~Dell’Elce, and G.~Kerschen, ``Analytic propagation of near-circular satellite orbits in the atmosphere of an oblate planet,''  {\em Celestial Mechanics and Dynamical Astronomy}, Vol.~123, Sept. 2015, pp.~85--103, 10.1007/s10569-015-9630-7.

\end{thebibliography}

\end{document}